**Rapid quench annealing of Er implanted Si for quantum networking applications**

Mark A. Hughes,[1*] Huan Liu,[4] Adam Brookfield,[2] Tianrui Wang,[3] Iain F. Crowe,[3] Yaping Dan[4]

[1]*School of Science, Engineering and Environment, University of Salford, Salford, M5 4WT, UK*
[2]*Department of Chemistry, University of Manchester, Oxford Road, Manchester, M13 9PL, UK*
[3]*Photon Science Institute and Department of Electrical and Electronic Engineering, University of Manchester, Manchester M13 9PL, UK*
[4]*University of Michigan-Shanghai Jiao Tong University Joint Institute, Shanghai Jiao Tong University, Shanghai 200240, China*
*m.a.hughes@salford.ac.uk

**Abstract**

Erbium implanted silicon (Er:Si) is a promising platform for quantum networking applications, but a major obstacle is the formation of multiple Er centres. We show that the previously identified cubic centre (Er-C) has $C_{2v}$ or lower symmetry. Using crystal field analysis of Er-C and other Er centres, and by comparison with extended X-ray absorption fine structure (EXAFS) measurements, we show that Er centres can be arranged in a sequence, ranging from entirely Si coordinated, through mixed Si and oxygen (O) coordination, to entirely O coordinated. G-factors calculated from our crystal field fitting closely match those determined by Zeeman splitting and electron paramagnetic resonance (EPR) measurements. We co-implanted Si with Er and O (each to a concentration of $10^{19}$ cm$^{-3}$). By increasing the quenching rate of the subsequent thermal anneal from ~100 °C/s to ~1000 °C/s, we change the dominant optically active centre, formed from $Er_2O_3$ clusters to the less energetically favourable Er-C centre with mixed Si and O coordination. Temperature dependent photoluminescence (PL) shows that $Er_2O_3$ clusters and Er-C centres have an O-related defect state at ~200 and 90 meV above the $^4I_{13/2}$ Er manifold, respectively. PL lifetime measurements show that the $Er_2O_3$ clusters and Er-C centres fall into two or three classes, characterised by different non-radiative PL decay rates. Our high quench rate annealing process could facilitate the formation of a single, optically active Er centre, which is preferable for quantum networking applications of Er:Si.

**Introduction**

The transmission of quantum information using photons at 1550 nm has many applications in quantum networking. For quantum key distribution (QKD), a quantum repeater is required, and entanglement swapping based quantum repeaters require a quantum memory. Atomic vapours and colour centres in diamond have been proposed for quantum memory applications, but atomic vapours aren't easily integrated with optical fibre or photonic waveguides, and have issues with atomic motions causing photon loss, while diamond colour centres necessitate a challenging conversion to telecoms wavelengths. The tendency of rare earths (REs) to have relatively large inhomogeneous linewidths and relatively narrow homogeneous linewidth makes them suitable for quantum memory protocols that involve spectral hole burning techniques. A notable protocol is Gradient Echo Memory (GEM) which uses a varying electric field, parallel to the photon propagation direction, applied to a high optical depth RE ensemble with a narrow "burnt in" absorption band. In principle, GEM efficiencies can reach ~100%, and it boasts the highest reported storage efficiency for a RE quantum memory of 69% using free space $Pr:Y_2SiO_5$ [1].

   With Er:Si, we can take advantage of well-established Si integrated circuit (IC) fabrication methods, as well as the well-established, highly repeatable and scalable networks that can be obtained using Si photonics, and the ability to integrate or interface with other Si based quantum computing platforms, such as donor, quantum dot, superconducting or photonic. A number of recent advances have been made in Er:Si, including Er implanted Si waveguides with a homogeneous linewidth of 9 kHz by Gritsch *et al* [2], our own reported spin $T_2$ time ~10 μs in Er implanted $^{nat}$Si [3], which has now been extended to the ~ms range in Er implanted $^{28}$Si [4], spin-resolved excitation of single Er centres with <0.1 GHz spectral diffusion linewidth [5], and our report of the first coupling with superconducting circuits [6].

An Er:Si waveguide based GEM memory could potentially improve efficiency over Er doped transparent bulk crystals by having uniform mode confinement along the beam propagation axis and precise control of the E-field gradient from the ability to precisely pattern electrodes on Si. This same quantum memory could be used in photonic quantum computers for synchronisation.

The Si T-centre has received attention because of a proposal from Photonic Inc. to use the manipulation of electron spins and spin dependent optical transitions from single T-centres in silicon optical cavities. Single photons from the T-centres are then interfered on a beam splitter to produce a maximally entangled Bell Pair spin state using the Barrett-Kok protocol [7]. This device could then be used as a resource for building fault tolerant quantum computers as well as quantum networking protocols including encoding based quantum repeaters [8]. Er:Si possesses many of the same attributes of the T-centre, which make it applicable to a telecoms Barrett-Kok protocol resource device: both have ~ms electron spin $T_2$ times, spin dependent telecoms transitions, nuclear spin $T_2$ times on the order of seconds; while the nuclear spin $T_2$ time of $^{167}$Er:Si has yet to be determined experimentally, it is likely to be similar to the 1.3 s determined for $^{167}$Er:Y$_2$SiO$_5$. However, the main disadvantage of Er:Si for this application is its ~ms spontaneous emission lifetime vs ~μs for the T-centre, which affects the indistinguishability of the interfered photons. However, a Purcell factor of >10$^5$ has been demonstrated in silicon photonic crystals [9], which would allow this to be mitigated, as an example of this possibility, Er implanted CaWO$_4$ coupled to a Si photonic cavity gave a Hong-Ou-Mandel (HOM) visibility of 80% [10].

Superconducting QCs could be permitted to function as larger scale, distributed systems over an optical fibre network with a Er:Si based microwave-to-optical quantum transducer. The large size of superconducting qubits means that coupling to an Er ensemble is desirable. Therefore, quantum memory based on Er ensembles in Si photonic waveguides and individual Er emitters in Barret-Kok protocol devices could enable a suite of quantum networking devices that are used in a future quantum internet.

When Er is implanted into Si, co-implantation or doping with another impurity, usually O, is required to observe strong luminescence by indirect, above bandgap, excitation, along with narrow EPR lines and n-type conductivity. At least six different spin centres have been identified by EPR [11, 12], along with numerous different luminescence centres, depending on the processing conditions [2, 13-15].

The ~100s nm optical path length of Er implanted Si substrates did not allow direct (resonant) excitation of Er until Gritsch et al observed directed excitation from Er:Si photonic waveguides (~mm optical path length) [2]. Although no co-doping was used, four Er emitting centres were observed. Arguably, the main issues with Er:Si are the number of different centres formed, the ability to control which centres are formed and an understanding of what those centres are. In order to fully realise the potential of Er:Si, the formation of these centres needs better control.

We have recently developed a 'deep cooling' process that follows thermal annealing. This involves using cryogenically cooled He gas to increase the post annealing quench rate by around an order of magnitude, compared to conventional rapid thermal annealing (RTA) and standard cooling cycles. We have previously used high concentrations of 7.5 × 10$^{20}$ cm$^{-3}$ Er and 2 × 10$^{21}$ cm$^{-3}$ O implanted Si combined with our 'deep cooling' process, to demonstrate reduced Er clustering and intense, room temperature (RT) luminescence with above bandgap excitation, which has a variety of photonic applications [16, 17]. We have also observed strong RT electroluminescence [16] and stimulated emission [18] from Er and O implanted Si LEDs, processed with 'deep cooling'. However, all our previous work on 'deep cooling' of Er/O implanted Si didn't have well resolved crystal field peaks due to the high Er and O concentrations used.

Here we apply the 'deep cooling' process to Si samples with much lower Er concentrations that are more applicable to quantum networking applications and by resolving crystal field splitting we show that the various Er centres can be represented on a sequence from entirely Si coordinated, through mixed O and Si coordination, to entirely O coordinated, with O coordination being energetically favourable. We show that in Er and O co-implanted Si, our 'deep cooling' technique can push the coordination towards greater Si coordination, which could be used in Er only implantation to obtain a single Si coordinated Er centre, which is essential for quantum networking application of Er:Si.

**Experimental**

**Sample fabrication**

Intrinsic (100) Si wafer, 500 μm thickness, supplied by Sigma-Aldrich, with a measured resistivity of 2 MΩcm was implanted with a chain of isotope specific $^{167}$Er implants up to 1.5 MeV and O up to 200 keV to give a flat concentration profile of ~$10^{19}$ cm$^{-3}$ for Er and O down to a depth of ~500 nm, see supplementary section S1. We chose this recipe because it has been shown by varying Er and O concentrations, that optimum luminescence intensity can be obtained from Er and O co-implanted silicon-on-insulator (SOI), with $10^{19}$ cm$^{-3}$ Er and $10^{19}$ cm$^{-3}$ O [19].

Annealing was carried out in a RTA, which had a peak cooling rate of ~100 °C/s, at 700 °C, 800 °C, 900 °C and 950 °C for 10 min, samples referred to as RTA 700, RTA 800, RTA 900, RTA 950, respectively and using 'deep cooling' (DC) at 900 °C and 950 °C for 10 min, samples referred to as DC 900 and DC 950, respectively. In order to perform 'deep cooling', we used a modified dilatometer (DIL 805A, TA Instruments), in which the samples, with a maximum width of 3 mm, were annealed at 900 °C and 950 °C for 10 min by induction heating at a pressure of 5 × 10$^{-4}$ mbar, followed by flushing with high-purity He (99.999%) gas cooled in LN2 (77 K) giving a peak cooling rate of ~1000 °C/s. A K-type thermocouple was used to monitor the temperature, which was controlled by adjusting the He flow rate.

PL spectra were obtained by placing the samples in a closed cycle He cryostat with a base temperature of 3.5 K. Samples were excited with a 450 nm, 50 mW laser diode powered by a laser diode driver that was electronically modulated by a function generator to give an overall fall time of ~1 μs. The PL was dispersed in a Bentham TMc300 monochromator and detected with an infra-red (IR) sensitive photomultiplier tube (PMT) and the signal was recovered with a SR830 lock-in amplifier. Transient PL measurements were taken using the same system used for spectral measurements, except the transient signal was captured with an oscilloscope with a 500 MHz bandwidth.

EPR measurements were taken on a Brucker EMX EPR spectrometer at a temperature of 5 K, with the magnetic field approximately parallel to the [110] crystal axis of the samples. The field modulation was 100 kHz, and the microwave frequency was 9.37 GHz. Raman measurements were performed on a Horiba labRAM evolution confocal Raman microscope with 488 or 633 nm laser excitation. Electrical conductivity measurements were carried out using a standard four-point probe.

**Crystal field analysis**

The Hamiltonian (H) of $Er^{3+}$ can be described as:

$$H = H_F + H_{CF} + H_{Ze} \quad (1)$$

Where $H_{CF}$ describes the electric field produced by the environment of the host crystal lattice surrounding the $Er^{3+}$ ion, and is given by the linear combination of spherical tensors of various ranks, $C_q^{(k)}$, and crystal field parameters (CFPs), $B_q^k$, as shown in Eq. 2:

$$H_{CF} = \sum_{k,q} B_q^k C_q^{(k)} \quad (2)$$

Only rank $k$ = 2, 4 and 6 are allowed, with q taking values of 0 to k, and taking into account the imaginary $C_q^{(k)}$, for q > 0, gives a total of 27 $C_q^{(k)}$ and associated $B_q^k$ CFPs. Each crystallographic point group has its own set of CFPs [20]. Details of the construction of $H_{CF}$ are given elsewhere [21]. Each $^{2S+1}L_J$ manifold will have its own $H_{CF}$ and the crystal field splitting is given by the eigenvalues of $H_{CF}$. To determine CFPs from experimentally derived crystal field splitting we used a least square algorithm that varied the $B_q^k$ values to minimize the difference between the observed splitting and the eigenvalues of $H_{CF}$.

$H_F$ accounts for the interactions that occur in a free RE ion. Each RE has its own set of $H_F$ parameters, and these vary little between hosts. For the fitting process, we didn't take into account $H_F$ but we used the $H_F$ parameters given by Carnall *et al.* for Er:$LaF_3$ [22] to calculate the absolute energy of excited state manifolds.

In Kramers-type REs, such as Er, with low symmetry, the crystal field splits each $^{2S+1}L_J$ manifold into J+½ degenerate crystal field doublets. The Zeeman interaction, $H_{Ze}$, which splits these doublets in a magnetic field is given in Eq. 3:

$$H_{Ze} = g_J \mu_B \mathbf{J} \cdot \mathbf{H} \quad (3)$$

Where $g_J$ is the Landé factor, $\mu_B$ is the Bohr magneton, **J** is the angular momentum operator, and **H** is the magnetic field strength [23]. $H_{Ze}$ is not considered in our crystal field fitting procedure.

**EPR *g* tensor calculation**

There are two sets of eigenvectors for each degenerate crystal field doublet: $|+\rangle$ and $|-\rangle$. The First order perturbation equations in Eq. 4 are used to determine the diagonal components of the g tensor, $g_x$, $g_y$, and $g_z$ [23].

$$g_x = 2g_J \langle +|J_x|-\rangle, \quad g_y = 2g_J \langle +|J_y|-\rangle, \quad g_z = 2g_J \langle +|J_z|+\rangle \quad (4)$$

Where $\mathbf{J}_x, \mathbf{J}_y, \mathbf{J}_z$ are the vector components of **J** such that $\mathbf{J}^2 = J_x^2 + J_y^2 + J_z^2$.

# Results and discussion

## Photoluminescence

Fig 1 a) shows the normalized PL spectra of the DC 950 sample at different modulation frequencies, measured at 3.5 K. Approximately fourteen relatively narrow peaks are identified, all with FWHM ~20 cm$^{-1}$, which is more than the maximum of eight peaks that could result from a single Er centre, indicating that the PL must originate from more than one Er centre. Using the fact that different Er centres could have different lifetimes, we increase the laser modulation frequency to reduce the PL intensity. In the ratio of low to high modulation frequency spectra, negative peaks correspond to peaks with shorter lifetimes. Using this technique, we can distinguish PL peaks arising from two different centres: a short lifetime centre (black arrows) and a long lifetime centre (red arrows). In the short and long lifetime centres the PL intensity decreased by a factor of ~4 and ~5, respectively, indicating that their difference in lifetime is only ~20%. In Fig. 1 b), we used the same technique but increased the temperature to 65 K; at this increased temperature, two peaks emerge on the short wavelength (higher energy) side of the most intense peak. These can be attributed to transitions from the first two thermally excited crystal field levels of the $^4I_{13/2}$ manifold to the lowest crystal field level of the $^4I_{15/2}$ ground state manifold (hot lines).

Fig. 1 c) shows the PL spectrum of the RTA 950 sample. Even though this sample was part of the same implanted wafer and had the same annealing temperature and time as the DC 950 sample, the PL spectra are markedly different: peaks are broader and less discernible than in DC 950, and the peaks that can be observed are at different wavelengths. The normalised spectra of the low and high modulation frequency are almost identical, as evidenced by the almost featureless spectrum of the ratio of these modulation frequencies. This suggests that all the peaks from RTA 950 are from the same Er centre. Integrated PL intensity was ~2× larger in DC 950 compared to RTA 950.

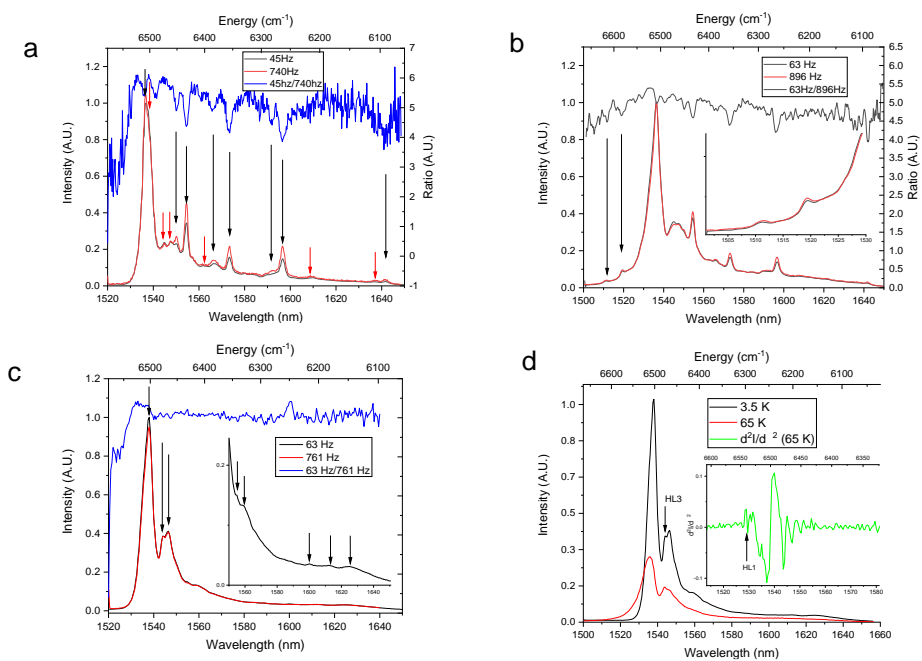

Fig 1 a) normalised PL from DC 950 at 3.5 K at slow and fast modulation frequencies, along with the ratio of slow to fast modulation spectra. b) normalised PL from DC 950 at 65 K at slow and fast modulation frequencies, along with the ratio of slow to fast modulation spectra. Inset shows a closeup of the hot lines. c) normalised PL from RTA 950 at 3.5 K at slow and fast modulation

*frequencies, along with the ratio of slow to fast modulation spectra. Inset shows a closeup of the long wavelength region. d) PL from RTA 950 at 3.5 and 65 K, a hot line (HL3) is indicated. Inset shows the 2nd derivative of the 65 K PL to resolve the first hot line (HL1), as indicated by the arrow.*

Fig. 1 d) shows the 3.5 K and 65 K spectra of the RTA 950 sample, with main peaks at 6501 cm$^{-1}$ and 6510 cm$^{-1}$, respectively; this thermally induced shift in peak position did not occur in DC 950. Thermal broadening of the emission peaks was also more apparent in RTA 950, than for DC 950. The intensity of the peak at 6476 cm$^{-1}$, marked HL3, increases with temperature, indicating that it is also a hot line. We have previously observed a hot line at 6472 cm$^{-1}$ in high concentration Er and O implanted Si, treated with our 'deep cooling' process[17]. Hot lines on the high energy side of the main peak can't be clearly observed due to thermal broadening of the most intense emission peak at 65 K, so to identify them we took the 2nd derivative of the PL spectrum, which can be used to enhance spectral resolution; in this case, emission lines correspond to negative peaks, i.e., for $d^2I/d\lambda^2 < 0$,[24, 25]. This procedure reveals a separate hot line (HL1) at 6538 cm$^{-1}$.

As can be seen in supplementary Fig. S2 a), the spectral position of emission lines in the PL of DC 900 and DC 950 are identical, although for DC 900 the short lifetime centre peaks are relatively weaker, and the long lifetime peaks relatively stronger than DC 950. Similarly, supplementary Fig. S1 b) shows that the RTA 900 and RTA 950 samples also have very similar spectra. This indicates that differences in the PL spectra between the DC and RTA samples are due to the quenching rate, rather than any possible differences in the annealing temperatures of the two samples. In both DC and RTA samples, the integrated PL intensity was ~2× higher for the 900 °C anneal than for the 950 °C anneal. Raman measurements in supplementary section S3 show that when annealing at temperatures of 700 and 800 °C by RTA, some residual amorphization and strain is evident, but when annealing at or above 900 °C (for both RTA and DC samples) no such residual amorphization or strain was evident.

**Inconsistencies of the cubic centre assignment**

In early work on Er:Si, Tang et al. reported a PL spectrum with five peaks when annealing at 900 °C for 30 min. These were attributed to a cubic (Er-C) centre on the basis that five peaks would be expected from an Er centre with cubic symmetry [26]. More peaks where observed with lower temperature anneals, which was attributed to the presence of a non-cubic Er centre [18]. In later, more extensive work, Przybylinska *et al.* observed the same five peaks of the cubic centre along with its first two hot lines in Er implanted float zone (FZ) Si, and the appearance of additional peaks in samples with higher O concentrations, which was attributed to the presence of non-cubic, O-coordinated centres [13]. If we compare the peak positions of our DC 950 short lifetime (DC 950 SL) centre to Er-C, as can be seen on the left side of Fig. 2, the most prominent PL peaks from our DC 950 SL peaks and the hot lines match very closely with Er-C (denoted 'Przybylinska Cubic'). There are three weak satellite peaks in the $^4I_{15/2}$ manifold of the DC 950 SL spectrum, which gives the total of eight peaks expected for a low symmetry site, that were not previously reported for Er-C. On the righthand side of Fig. 2, we also show the peaks from our RTA 950 sample, compared with our DC 950 long lifetime (DC 950 LL) spectrum and those from RTA 950, which shows that the first four DC 950 LL CF levels match those in RTA 950, indicating that the DC 950 LL peaks represent the remnant of the remaining RTA 950 centre.

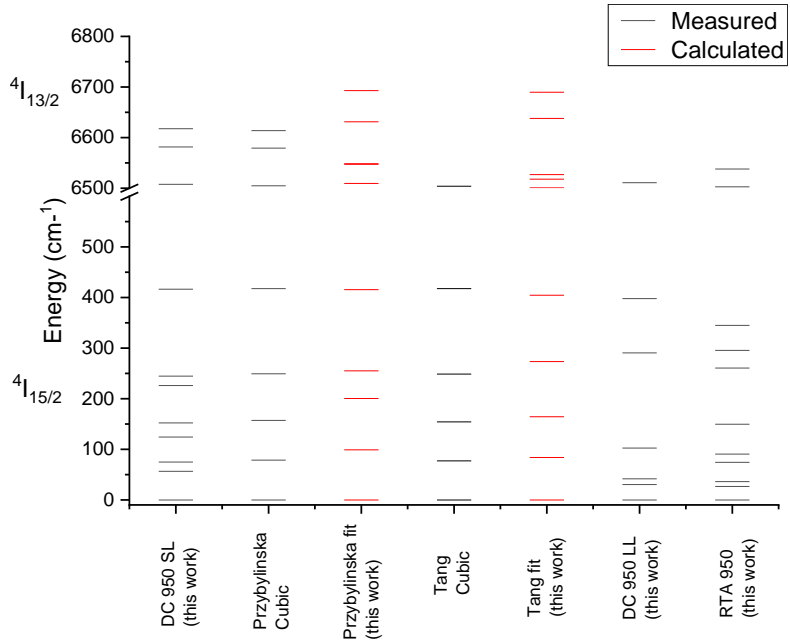

*Fig. 2 crystal field splitting of our RTA 950 and DC 950 samples, compared with supposed cubic crystal field splitting from the literature, Przbylinska [13] and Tang [26].*

If we examine the literature on Er:Si PL, the same five PL lines can be seen; however, the same satellite peaks that we observe in DC 950 SL can also be observed, but were not commented on [15, 27-32]. A wide variety of implantation and annealing conditions were used in these studies, yet all produce the same characteristic peaks.

A cubic centre can be fitted with four CFPs, two of which have a constant ratio to the other two, giving two adjustable parameters to describe the cubic crystal field. These two CFPs can be re-expressed so that relative splitting is described by a single parameter, x, and a linear scaling factor, W, which can be thought of as a stretch of the splitting, determined by x [33].

The irreducible representations, Γ, of the cubic group give the crystal field levels. As shown in Fig. 3, in a cubic field the $^4I_{15/2}$ manifold splits into two Kramers doublets ($Γ_6$ and $Γ_7$) and three $Γ_8$ quartets. A $Γ_a → Γ_b$ transition is subject to selection rules: the transition is allowed if the reduction of $Γ_K ⊗ Γ_b$ contains $Γ_a$, in cubic symmetry, $Γ_K = Γ_5$ [34]. This results in $Γ_6 → Γ_6$ and $Γ_7 → Γ_7$ transitions being forbidden.

Cubic CFPs have previously been fitted to the $^4I_{15/2}$ splitting of the Er-C centre [13, 35]. We have also carried out the fitting of the splitting reported by Tang[26] here, see Fig. 2, and find similar fitting parameters (x = 0.341, W = 0.970) with a root mean square deviation (RMSD) = 30.5 cm$^{-1}$. However, the predicted splitting of the $^4I_{13/2}$ manifold has no match to the observed hot lines. Also, given the selection rules, this presents a problem for the fit with x = 0.341, since $Γ_7$ is the ground state of the $^4I_{13/2}$ manifold, so only four transitions to the $^4I_{15/2}$ manifold are allowed. In fact, for all five transitions to be allowed, $Γ_8$ needs to be the ground state of the $^4I_{13/2}$ manifold, which only occurs for -1 < x < -0.4, but the splitting in this x range does not match the spectrum of Er-C. One possibility is that x is actually at the cross-over of $Γ_6$ and $Γ_7$ in the $^4I_{13/2}$ manifold at x = 0.65, because of the range of crystal field sites giving a range of x values for the Er-C centre there could be transitions from both $Γ_6$ and $Γ_7$ so all five transition could be observed. The problem with this is that small changes in the crystal field that could be expected for different processing conditions would change x slightly. This would change which transitions are observed, but the peaks of the Er-C centre are very stable and reproducible, the fit to the $^4I_{15/2}$ splitting is made significantly worse (RMSD = 82 cm$^{-1}$) and the splitting of the $^4I_{13/2}$ manifold cannot explain the observed hot lines. In fact, if we attempt to fit all the known splitting of Er-C ($^4I_{15/2}$ PL and $^4I_{13/2}$ hot lines), there is no acceptable fit to cubic CFPs. We can have a high confidence in the null fit result precisely because the peaks of the Er-C centre are so stable and reproducible, and there are only two degrees of freedom for the cubic fit, so it is relatively easy to search the entire parameter space. Along with the consistent observation of three satellite peaks and the fact that five peaks can only be observed for a very limited set of cubic crystal fields, this leads us to propose with confidence that Er-C is not, in fact, a cubic centre; however, for consistency with the literature we refer to this centre as Er-C.

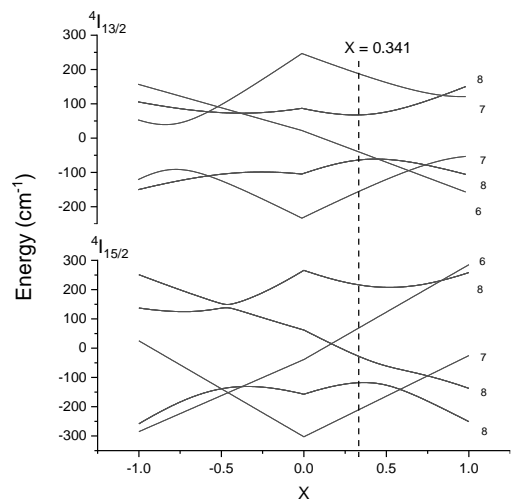

*Fig. 3 Cubic crystal field splitting for $^4I_{15/2}$ and $^4I_{13/2}$ manifolds*

**Crystal field analysis**

Crystal field analysis is a powerful tool in determining which RE centres are present from their optical spectra. Typically, this is done with RE doped transparent bulk crystals which allows observation of the stark splitting of many manifolds, by absorption measurements giving, perhaps, hundreds of experimental energy levels to fit both crystal field and free ion parameters. For Er:Si, much more care must be taken in performing and interpreting crystal field analysis since, from indirect PL measurements, we only have the splitting of the $^4I_{15/2}$ manifold, plus a few (hot line) levels from the $^4I_{13/2}$ manifold. More recently the complete $^4I_{13/2}$ splitting of a few Er centres from direct PL of Er:Si waveguides [2] and optically modulated magnetic resonance (OMMR) [36] were added. The significance of a good fit to a single manifold should be downplayed, especially considering the previous confusion over a cubic fit to the Er-C centre. Here, we use the many centres observed in Er:Si to our advantage by assuming that all, or at least a subset, of these centres can be placed in a sequence that represents some monotonic change in their local environment. A monotonic change in local environment, such as a shift in primitive cell dimension or shift in coordination from Si to O, for example, would be expected to give a monotonic shift in fitted CFPs [21]. To put the centres in a sequence we look for trends in their splitting, and then look for trends in the fitted CFPs, which would be expected for a monotonic change in local environment. This effectively gives more data points for the analysis and validates one fit against another. We also calculate g-factors from the fits and compare to EPR and Zeeman measurements for further validations.

In order to determine which symmetry group CFPs to fit, we can employ a descending symmetries technique [37]. Starting with the cubic $O_h$ symmetry of the Si host, which is ruled out with a high degree of confidence, the eight lines observed in the Er-C $^4I_{15/2}$ manifold means it must have tetragonal or lower symmetry. The next lowest symmetry progressing down the symmetry subgroups [38] is tetragonal $D_{2d}$ [21]; however, no good fit for the DC 950 SL lines could be found for tetragonal CFPs. The main issue is the large separation of the first two crystal field levels in the $^4I_{13/2}$ manifold. Since tetragonal symmetry is a distortion of cubic symmetry, unless the tetragonal distortion is very large, there will be a lot of similarity to the splitting seen in Fig. 3. We attempted to fit hexagonal and trigonal CFPs, which were not as good as orthorhombic, ruling out the hexagonal branch, or the crystallographic subgroups. The next lowest symmetry in this series is orthorhombic $C_{2v}$ symmetry, which has nine CFPs and is often employed for the crystal field analysis in situations where the symmetry is uncertain [37]. With this number of degrees of freedom in the fit, care must be taken to avoid local minima, so we randomly adjusted starting parameters to help find the global minimum. An additional issue is that there could be unresolved peaks, or peaks from a different centre in any spectrum, although our method of arranging spectra in a sequence should help identify the presence of any such peaks.

From the first resonant excitation of Er:Si waveguides, Gritsch et al identified sites A, B, P and O [2]. Sites A and B are of particular interest since they have not been observed before. Site P was attributed to precipitates and site O was attributed to an indirectly excitable centre. The Gritsch A and B centres are consistent with our previous indirect measurement of the $^4I_{13/2}$ splitting using OMMR [36], see supplementary section S4.

The measured splitting of Gritsch site A and B, and DC 950 SL can be arranged in a sequence, as shown in Fig. 4, along with the fitting of these with orthorhombic CFPs. All of these were reasonably well reproduced, with Gritsch site A being a particularly good fit. Measured Gritsch sites A and B exhibited only six levels in the $^4I_{15/2}$ manifold, which could be due to the long wavelength cutoff of the detection system used in their experiments. By allowing two levels in the $^4I_{15/2}$ to vary freely, the fits predict the position of the two longest wavelength PL peaks. Our proposed RTA 950 crystal field splitting is also shown. The first excited Stark level of the $^4I_{13/2}$ manifold was determined from the 2$^{nd}$ derivative of the 65 K PL spectrum in Fig. 1 d) and corresponds to hot line, HL1. The hot line, HL2 transition would approximately overlap with the main PL peak, which explains why it was not observed, and could explain the shift and excessive thermal broadening of the main PL peak, as shown in Fig 1 d). The hot line, HL3 transition corresponds to the 6476 cm$^{-1}$ peak in Fig. 1 d), and is consistent with hot line, HL1.

The RTA 950 splitting could not be put in the same sequence as the Gritsch A, B and DC 950 SL levels, but it does closely match the Stark splitting of $Er_2O_3$ [39], which has an almost identical splitting to Er doped $Y_2O_3$ [40] in $C_2$ sites, and in both of these materials, Er is 6 fold coordinated to O. The missing level at ~155 cm$^{-1}$ may not have been resolved because of inadequate spectral resolution. Additional peaks at 295 and 345 cm$^{-1}$ can be attributed to Er in different sites, which was also observed in Er doped $Y_2O_3$ [40]. Because of the strong similarity, we propose that the emission peaks in the RTA 950 spectrum originate from the same six-fold O coordinated Er centre, with $C_2$ symmetry as that observed in $Er_2O_3$ and Er:$Y_2O_3$. The splitting of Er:$Y_2O_3$ was previously fitted to the 14 CFPs of the $C_2$ point group [40], but for consistency with our fitting to Gritsch A, B and DC 950, we fitted with the 9 CFPs of $C_{2v}$, of which $C_2$ is a subgroup.

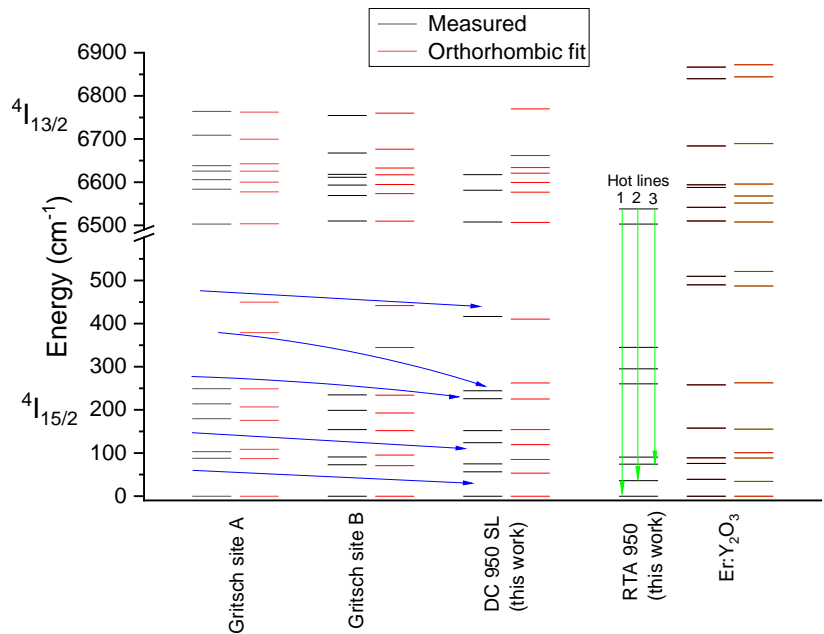

*Fig. 4 Measured crystal field splitting and orthorhombic fit to Gritsch site A and B [2] DC 950 SL, RTA 950 (this work) and Er:$Y_2O_3$ [40]. Blue arrows indicate the trend in monotonic shift, green arrows indicate the proposed hot line transitions for RTA 950 (HL1,2,3)*

Fig. 5 shows the orthorhombic CFPs from the fits shown in Fig. 4. Since the $C_{2v}$ symmetry can be thought of as an orthorhombic distortion to a tetragonal $C_{4v}$ field, we separate the $C_{2v}$ CFPs that are common to $C_{4v}$. Fig. 5 shows a monotonic trend in all CFPs for the sequence Gritsch A, B and DC 950, as would be expected for a monotonic shift in local environment. It also shows that the $C_{2v}$ fit to Er:$Y_2O_3$, and by extension RTA 950, does not follow the same sequence.

The RTA 950 centre is the six-fold O coordinated Er centre in $Er_2O_3$. However, in order to form this centre, an O:Er ratio of ~6 is required, but RTA 950 only has an O:Er ~1. It is possible for O to diffuse to the surface during annealing, but this would be limited a depth of ~10 μm [41-43]. Given the background O concentration of our FZ wafer is ~$10^{15}$ cm$^{-3}$, the amount of diffused O available is at most ~$10^{12}$ cm$^{-2}$, compared with the implanted Er dose of 4.18×$10^{14}$ cm$^{-2}$. So, either ~1/6 of the implanted Er is used to form individual six-fold O coordinated Er centres, or ~2/3 of the implanted Er is used to form $Er_2O_3$ clusters. $Er_2O_3$ cluster formation is the more likely explanation, otherwise we would have to explain why the remaining 5/6 of the implanted Er formed some indirect excitation inactive centre, such as Gritsch A and B, rather than Er-C, which is readily formed in Er:Si.

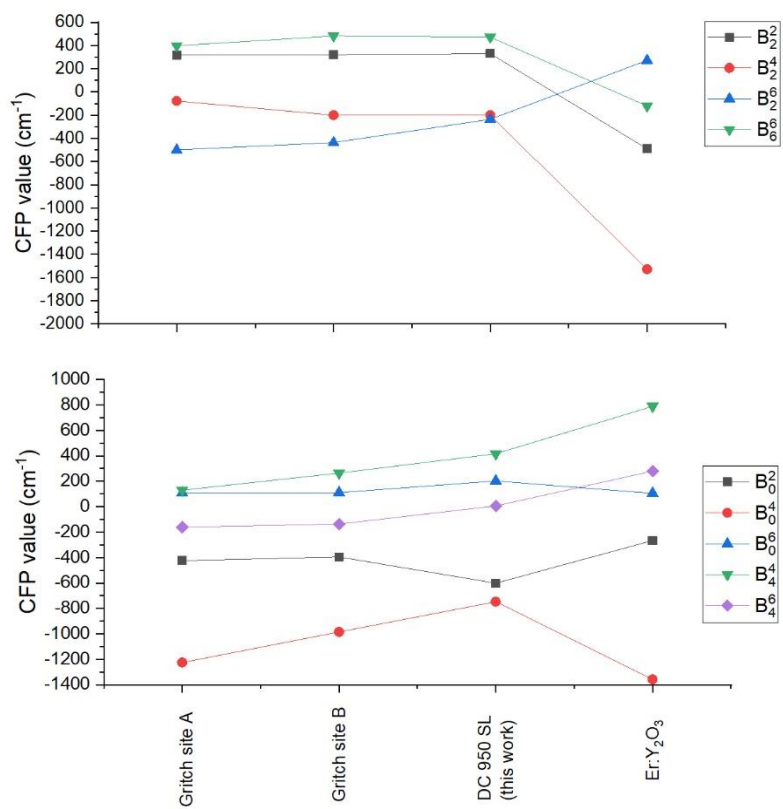

*Fig. 5 $C_{2v}$ Orthorhombic CFPs from fits shown in Fig. 4. The bottom panel shows the CFPs common to $C_{4v}$ symmetry, whereas the top panel shows those unique to $C_{2v}$ symmetry.*

**Comparison to EXAFS**

A range of Er PL centres associated with co-doping with O [13, 15, 44] that are yet to be classified in our sequence in Fig. 4 are referred to as "unclassified". Extended X-ray absorption fine structure (EXAFS) measurements have been performed on Er:Si samples with a range of implantation and annealing conditions. These are summarised in Table 1, along with the PL assignment from those, or similar, Er:Si samples. We can put the reported coordination of Er in a sequence: Si only coordination (12- or 6-fold), mixed Si and O coordination, O only coordination (4- to 6-fold). In the associated PL assignment column, we identify both the characteristic Er-C centre, along with those categorised as "unclassified". This associated PL can be summarised as: very weak Er-C PL from Si only coordination, clear, strong Er-C PL from mixed Si and O coordination and mixed Er-C and "unclassified" PL from O only coordination. We propose that the Er-C centre has mixed coordination, the Gritsch A and B centres have entirely, or close to entirely, Si coordination. The reason Gritsch A and B centres are not observed by indirect excitation is because the O is needed to create a defect level to allow indirect excitation. The $Er_2O_3$ clusters in RTA 950, and the as yet unclassified Er:Si PL [13, 36, 44] have 4- to 6-fold O coordination.

*Table 1 summary of Er:Si EXAFS measurements and their associated PL assignment from the literature. ‡ is shorthand for 450 °C for 30 min, 620 °C for 3 hr. †Where PL was not reported in the EXAFS publication, PL ref gives the closest sample from the literature with a PL measurement. "Unclear" means that the PL spectrum was too featureless or noisy to identify a PL centre.*

| O coord | Si coord | Implant | wafer | anneal | ref | PL | PL ref† |
|---|---|---|---|---|---|---|---|
| 5.5 | - | $10^{19}$ Er, $10^{20}$ O | CZ | 900 °C 12 hr | [45] | | |
| 5.9 | - | $10^{19}$ Er, $10^{20}$ O | CZ | 900 °C, 30s | [45] | unclassified | $10^{19}$ Er, $10^{20}$ O; 900 °C, 30s [46] |
| | 12 | $5\times10^{17}$ Er | FZ | 927 °C, 30 min | [47] | Er-C v weak | $5\times10^{17}$ Er; 900 °C 30 min [13] |
| 6 | | $5\times10^{17}$ Er | CZ | 927 °C, 30 min | [47] | | |
| 3 | 6 | $10^{19}$ Er, $3\times10^{19}$ O | | ‡, 900 °C, 30 s | [48] | Er-C + weak unclassified | |
| 5.1 ±0.5 | - | $10^{19}$ Er, $10^{20}$ O | | ‡, 900 °C, 30 s | [48] | Er-C + unclassified | |
| 4 ±1 | - | $10^{19}$ Er, $10^{20}$ O | | ‡, 900 °C, 12 hr | [48] | Unclear | |
| | 6 ±2 | $10^{19}$ Er, $10^{20}$ O | | 450 °C, 30 min | [49] | Unclear | |
| 3 ±2 | 4.4±0.6 | $10^{19}$ Er, $10^{20}$ O | | ‡ | [49] | Er-C + unclassified | |
| 5.1±0.5 | | $10^{19}$ Er, $10^{20}$ O | | ‡, 900 °C, 30 s | [49] | Er-C+ unclassified | |
| 5.0±0.5 | | $10^{19}$ Er, $10^{20}$ O | | ‡, 900 °C, 12 hr | [49] | Unclear | |

The associated PL can be explained by the sensitivity limit of EXAFS: the 12 Si only coordination FZ sample has very weak Er-C PL because only a few Er-C centres, below the EXAFS sensitivity limit, can be formed from the residual O impurities in FZ Si. The majority of the Er is Si only coordinated, the PL for which is only activated by direct (resonant) excitation. Similarly, for the O only coordinated Er, which has mixed Er-C and unclassified PL emission lines, a fraction of Er has the mixed Er-C coordination, while the majority has the O only, unclassified Er centres.

     Both EXAFS measurements [47] and DFT calculations [50] indicated the preference for Er to coordinate with O impurities over Si, which suggests that O coordination is energetically favourable to Si. We propose that the reason 'deep cooling' reduces O coordination is that at the annealing temperatures used, O coordination is less energetically favourable and, during annealing the Er is preferentially coordinated with Si. With a slow quench there is sufficient time at lower temperatures (where atomic rearrangement can still occur, but O coordination is favourable) for the higher O coordinated Er centres to form, whereas with a rapid quench there is insufficient time for Er to coordinate with O. This proposed mechanism has some analogies to the quenching of a molten glass, where a rapid quench can form the less energetically favourable amorphous state, and a slow

quench can form the more energetically favourable crystalline state. The proposed Er centres are summarised in Table 2.

*Table 2 Summary of proposed centres formed in Er implanted Si. †Indirect via O-related defect state*

| Centre | O coordination | Si coordination | Symmetry | PL excitation | Energetic favourability |
|---|---|---|---|---|---|
| Gritsch A | 0 | 6-12 | $C_{2v}$, or lower | Resonant only | Least favourable |
| Gritsch B | 1-2 | 6-12 | $C_{2v}$, or lower | Resonant only | |
| Er-C | 2-6 | 2-6 | $C_{2v}$, or lower | † | |
| Unclassified | 4-6 | 0-2 | unknown | † | |
| RTA 950 | 6 | 0 | $C_{2v}$, or lower | † | Most favourable |

**g-factor calculation**

In Er:Si, particular processing conditions are required to obtain EPR active centres. Supplementary section S5 shows that no EPR signal could be obtained for DC 950, whereas for $10^{19}$ cm$^{-3}$ Er and $10^{20}$ cm$^{-3}$ O, a strong EPR signal is observed. If an EPR signal cannot be measured from optically active Er:Si, given a sensitive enough method, Zeeman measurements can still be performed. Yang et al used Zeeman measurements from a $1 \times 10^{17}$ cm$^{-3}$ $^{167}$Er and $1 \times 10^{18}$ cm$^{-3}$ O co-implanted FinFET transistor, annealed at 700 °C, to obtain the g-tensor, hyperfine A-tensor, and their associated Euler angles for the CF ground states of the $^4I_{15/2}$ and $^4I_{13/2}$ manifolds.

In Table 3 we show the principal values of the g-tensors calculated from the CFPs obtained from our fitting. There is a very good agreement between both the $^4I_{15/2}$ and $^4I_{13/2}$ calculated g-tensors of the Gritsch B centre and those measured by Yang. There is also good agreement between the g-tensor calculated from our $C_{2v}$ fit to Er:Y$_2$O$_3$ and the OEr-3 centre, derived from EPR by Carey et al [51]. We can accurately predict the measured g-factor of Er:Y$_2$O$_3$ from our $C_{2v}$ fit to its Stark levels, which serves as a good test case of our method.

Yang et al concluded their Er centre had monoclinic $C_1$ symmetry based on misalignment between the $^4I_{15/2}$ and $^4I_{13/2}$ g-tensors. This is consistent with our finding of $C_{2v}$ symmetry since our method of descending symmetries will give the highest symmetry capable of reasonably explaining the observed splitting added to the fact that $C_1$ symmetry is obtained by applying a distortion to $C_{2v}$ symmetry.

*Table 3 Principle g-tensors calculated from the CFPs in Fig. 5 along with g-tensors measured by Zeeman [52] and EPR [51]*

| Centre | g-tensor | | | | | | Symmetry | Ref |
|---|---|---|---|---|---|---|---|---|
| | $^4I_{15/2}$ | | | $^4I_{13/2}$ | | | | |
| | $g_x$ | $g_y$ | $g_z$ | $g_x$ | $g_y$ | $g_z$ | | |
| | Calculated | | | | | | | |
| Gritsch A | 0.40 | 1.36 | 16.36 | 0.16 | 0.33 | 14.06 | $C_{2V}$ | This work |
| Gritsch B | 0.72 | 2.91 | 14.98 | 0.32 | 0.70 | 13.75 | $C_{2V}$ | This work |
| Er-C | 1.49 | 3.04 | 14.80 | 0.38 | 0.48 | 13.99 | $C_{2V}$ | This work |
| Er:Y$_2$O$_3$ | 1.02 | 3.96 | 11.83 | 0.86 | 1.14 | 12.61 | $C_{2v}$ | This work |
| | Measured | | | | | | | |
| Yang | 0.55 ± 0.19 | 2.38 ± 0.18 | 14.846 ± 0.028 | 0.16 ± 0.16 | 0.59 ± 0.17 | 13.1 ± 0.023 | $C_1$ | Zeeman [52] |
| Carey OEr-3 | 1.09 | 5.05 | 12.78 | | | | $C_1$ | EPR [51] |
| Er:Y$_2$O$_3$ | 1.645 | 4.892 | 12.314 | | | | $C_2$ | EPR [53] |

**Indirect excitation mechanism**

The indirect excitation mechanism for Er:Si is thought to involve transfer of energy, from excitons generated by above bandgap excitation, to the optically active Er site, via an intermediate defect state, associated with the presence of an Er-O centre in the Si lattice. Quenching of the excited Er is thought to occur either by Auger quenching (via impurity-Auger and/or exciton-electron-Auger recombination) [28], with an associated activation energy $E_{AQ}$, or via back-transfer to the intermediate defect level by exciton dissociation and formation of free carriers [54], with an associated activation energy $E_{BT}$ [55], as illustrated in Fig. 6 a).

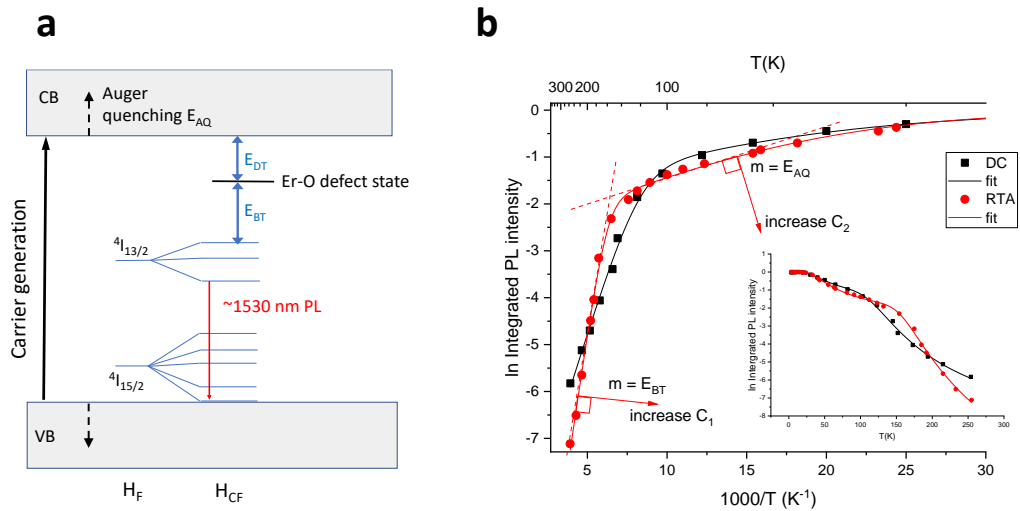

*Fig. 6 a) Standard energy level model for Er:Si illustrating (not to scale) the free ion splitting, $H_F$, and crystal field splitting, $H_{CF}$, of Er along with the defect state and possible quenching mechanisms. b) Arrhenius plot of the temperature dependence of the Er$^{3+}$ integrated PL intensity with 450 nm excitation for DC 950 (black) and RTA 950 (red). Lines describe fits using Eq. 5. Inset is the normalised intensity Vs temperature data for the same samples*

Fig. 6 b) shows the Arrhenius plot of the temperature dependence of the integrated PL intensity from DC 950 and RTA 950. The general dependence is typical for that observed from Er:Si [13, 56-59] with a shallow slope at low temperature which transitions to a steeper slope at higher temperatures. This has been attributed to Auger quenching (at lower T) and (thermally assisted) back transfer (at higher T). These thermal quenching mechanisms are represented in Eq. 5 with $E_{BT}$, $E_{AQ}$ and their associated pre-factors ($C_1$, $C_2$). We fitted the temperature dependence with Eq. 5, as shown in Fig. 6 b), and the fit parameters are shown in Table 4. $E_{AQ}$ is similar to that observed previously [60], while $E_{BT}$ was previously reported to be 150 meV [55]. We find $E_{BT}$ is 200 meV and 90 meV for RTA 950 and DC 950, respectively. $E_{BT}$ is of importance because it indicates the energy of the defect state, relative to the first excited state of the Er centre. In this case, this would place the defect state at 160 and 270 meV below the conduction band edge, for RTA 950 and DC 950, respectively. Since this defect state is assumed to be related to O impurities around the Er centre, the difference in $E_{BT}$ is explained by the different O coordination expected for RTA 950 and DC 950, and implies that increased O coordination pushes the defect level towards the conduction band edge. The fact that $C_2 \ll C_1$ indicates that Auger is mostly ineffective in quenching, compared with the back transfer process.

$$I(T) = I_0 \frac{1}{1 + C_1 e^{(-E_{BT}/k_B T)} + C_2 e^{(-E_{AQ}/k_B T)}} \quad (5)$$

Table 4 Fitting parameters for temperature dependence of integrated $Er^{3+}$ PL intensity

|  | RTA 950 | DC 950 |
| --- | --- | --- |
| $C_1$ | 1.16E7 ± 3E6 | 25584 ± 5500 |
| $E_{BT}$ (meV) | 198.5 ± 4.8 | 93.3 ± 3.4 |
| $C_2$ | 13.5 ± 1.5 | 5.8 ± 1.8 |
| $E_{AQ}$ (meV) | 12.5 ± 0.8 | 9.9 ± 1.9 |

The n-type conductivity of Er:Si has been related to the presence of the Er-O defect state [61]. However, despite the significant difference in the position of the defect state for DC 950 and RTA 950 annealing, derived from Fig. 6 b), resistivity measurements in supplementary section S6 show no significant difference between the DC and RTA annealed samples. Additionally, we observed an increase in resistivity, by a factor of three, when the temperature for RTA annealing increased from 700 to 900 °C. Since crystallinity is not fully recovered for samples annealed at lower temperatures, as confirmed by Raman measurements of the 700 and 800 °C RTA samples, this suggests that electrical conductivity may in fact be facilitated by defects, such as O and dislocations, in the crystal lattice, rather than via the optically active Er-O centre, in a fully recrystallized lattice. Any conductivity, either due to O centres, dislocations or the Er-O defect, is unfavourable for quantum networking applications because free carriers could increase the propagation loss of Er:Si waveguides, via free carrier absorption (FCA).

**Transient PL**

Fig. 7 a) and b) show the fluorescence decay profiles associated with the principal emission peak for the RTA 950 and DC 950 samples, respectively. Neither can be explained by single exponential decays. DC 950 appears to be bi-exponential, as evidenced by the lack of improved fitting accuracy when increasing the number of exponential decays in the fit from two to three; $R^2$ values for single (0.875), bi- (0.916), and tri-exponential (0.916) fits, reveals the third exponential to be redundant in improving the fit accuracy. The RTA 950 data is less clear, but by increasing the number of exponential decays in the fit from one to four, revealed no improvement in $R^2$ beyond a tri-exponential fit; with the values for single (0.915), bi- (0.958), and tri-exponentials (0.979) indicating that the decay is most likely tri-exponential in nature.

The 273 µs and 2197 µs lifetime components of DC 950 do not represent the short and long lifetime components observed in Fig 1 a) as these only differed by ~20%. Measurement of the PL decay profiles was limited to the detection range of the principal (most intense) Er-related emission peak (1532-1542 nm), which we already identified as being associated with the DC 950 SL centre. The 2197 µs component is close to the expected radiative lifetime of Er, whereas the 273 µs component, although obviously from the same emitting Er centres, with the same local environment, is suggestive of a faster (possibly nonradiative) decay pathway (e.g., by being in close proximity to a defect)

For RTA 950, Fig 1 c) indicated no spectral dependence of lifetime, implying the same, single emitting centre with the same crystal field local environment was responsible for all peaks. However, from the PL decay data, it is clear that there are three different classes of this centre, with different nonradiative decay rates, all of which are significantly higher (lifetimes shorter) than those in the DC 950 sample.

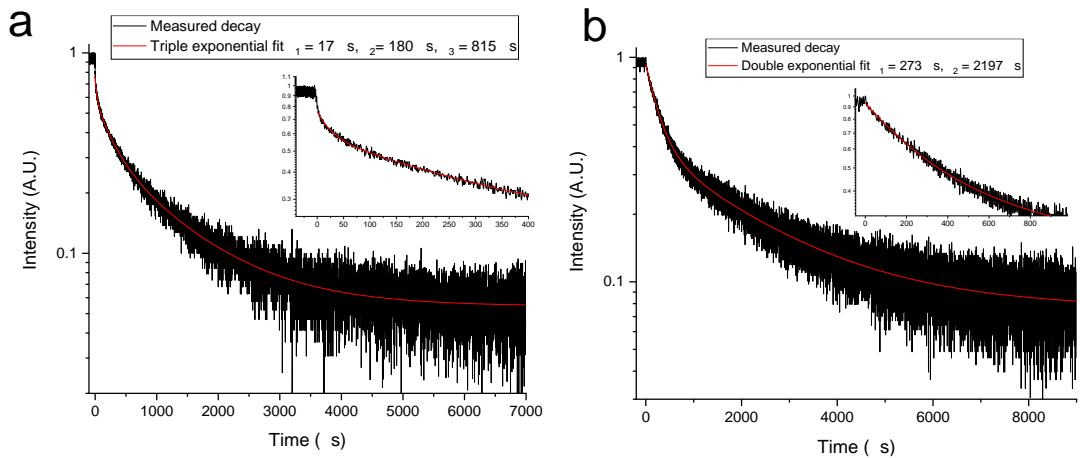

Fig. 7 Log-linear decay profiles of PL collected from ~1532 nm to 1542 nm at 3.5 K with 450 nm excitation for a) RTA 950 b) DC 950. Insets show close-ups of the decay transients in the short time window, where the PL intensity has decayed to ~ 1/e.

**Conclusions**

By increasing the quenching rate of the thermal anneal of Er and O co-implanted Si, using our 'deep cooling' method we change the dominant Er centre from $Er_2O_3$ clusters to a less energetically favourable centre with mixed Si and O coordination. We use a variable modulation frequency method to separate peaks belonging to different Er centres in Er:Si with different PL lifetimes. PL decay profiles show that these centres have multiple classes with different non-radiative decay rates.

We show that the spectra of various Er centres can be arranged in a sequence. By fitting CFPs we validate the existence of this sequence and show that the symmetry of these centres is $C_{2v}$, or lower. By calculating g-factors from the fitted CFPs, we validate the fit. From comparisons to EXAFS measurements, we propose that all Er centres in Er:Si can be arranged in a sequence, ranging from entirely Si coordinated, through mixed Si and oxygen (O) coordination, to entirely O coordinated. Er centres with higher O coordination have an associated O-related defect state which allows indirect excitation.

Future quantum networking applications of Er:Si will require the implantation of Er into high purity SOI; however, studies to date indicate that this results in the formation of four different Er centres. O-coordination of Er is energetically favourable, and a low concentration of O is likely to be unavoidable in SOI, as a result of O diffusion into the thin Si device layer, during the thermal oxidation step of SOI fabrication. We propose that the best case for quantum networking applications is the formation of only the 'Gritsch A' (i.e., Si-coordinated Er) centre, since this would avoid excess decoherence and propagation loss associated with other centres, in particular, those with higher O coordination and an O related defect state. The 'deep cooling' process we have devised can push the Er centre towards this preferable higher Si coordination state, despite it being energetically less favourable under more conventional thermal annealing cycles. This is likely to be helpful in obtaining only the Gritsch A centre for quantum networking applications of Er:Si.

# Supplementary information
# Rapid quench annealing of Er implanted Si for quantum networking applications

Mark A. Hughes,[1*] Huan Liu,[4] Adam Brookfield,[2] Tianrui Wang,[3] Iain F. Crowe,[3] Yaping Dan[4]

[1]School of Science, Engineering and Environment, University of Salford, Salford, M5 4WT, UK
[2]Department of Chemistry, University of Manchester, Oxford Road, Manchester, M13 9PL, UK
[3]Photon Science Institute and Department of Electrical and Electronic Engineering, University of Manchester, Manchester M13 9PL, UK
[4]University of Michigan-Shanghai Jiao Tong University Joint Institute, Shanghai Jiao Tong University, Shanghai 200240, China
*m.a.hughes@salford.ac.uk

**Section S1 Implantation simulation**

Undoped 2" wafers from Sigma-Aldrich were implanted at room temperature with a 7° tilt angle. with a chain of $^{167}$Er and O implants. The simulation of the implants using the SRIM Monte Carlo simulation code is shown in Fig. S1.

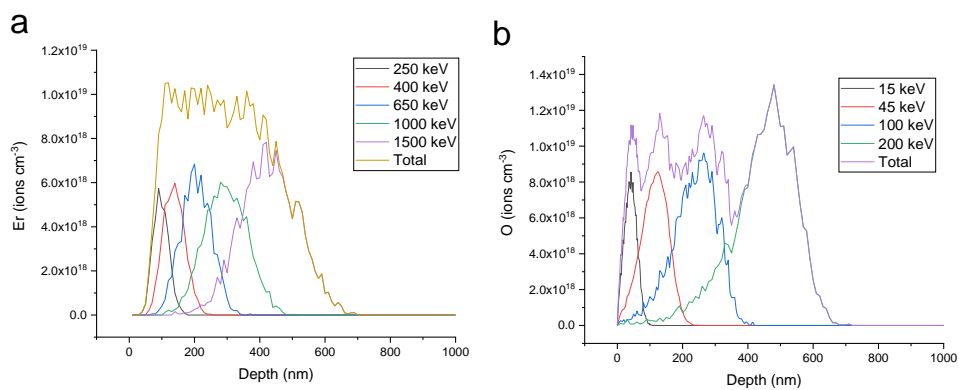

Fig. S1 a) SRIM simulation of the implant profile for $^{167}$Er with a simulated peak concentration of $10^{19}$ cm$^{-3}$ and a total areal dose of $4.18\times10^{14}$ cm$^{-2}$. b) Simulated implant profile for O with a simulated peak concentration of $\sim 10^{19}$ cm$^{-3}$ and a total areal dose of $5.5\times10^{14}$ cm$^{-2}$.

**Section S2 Comparison of PL from RTA and DC annealing over a range of annealing temperatures**

Fig. S2 a and b shows the PL spectra from using the DC and RTA annealing techniques at temperatures of 900 and 950 °C.

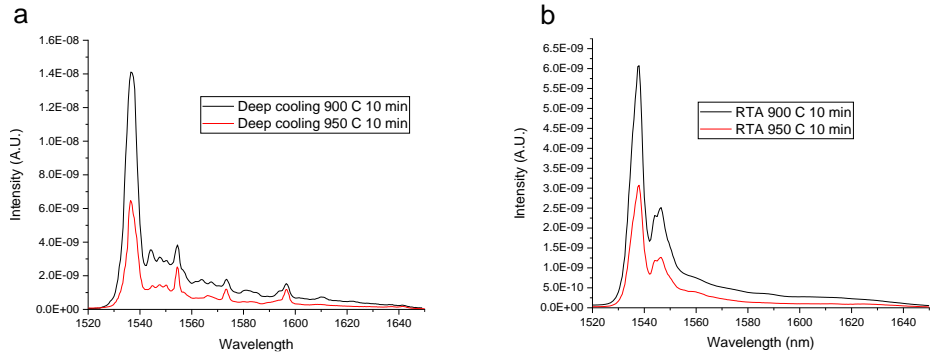

*Fig. S2 a PL spectra of $10^{19}$ cm$^{-3}$ Er, $10^{19}$ cm$^{-3}$ O co-implanted Si annealed by the DC process at temperatures of 900 and 950 °C for 10 min. b PL spectra of $10^{19}$ cm$^{-3}$ Er, $10^{19}$ cm$^{-3}$ O co-implanted Si annealed by RTA at temperatures of 900 and 950 °C for 10 min. For all spectra the excitation wavelength was 450 nm and the temperature was 3.5 K.*

**Section S5 Raman measurements**

Raman measurements were performed on a Horiba labRAM evolution confocal Raman microscope with a 1800 line/mm grating and using 488 or 633 nm laser excitation, which gave penetration depths of 490 nm and 3 µm, respectively, which effectively probed different depths of the sample. Raman spectra were taken at three different positions and fitted with Gaussian-Lorentzian peaks to obtain average peak positions and widths along with a standard error. Fig. S7 a) shows the first order optical vibration mode of Si at ~520 cm$^{-1}$ of the unimplanted wafer and $10^{19}$ cm$^{-3}$ Er, $10^{19}$ cm$^{-3}$ O co-implanted Si annealed by RTA and DC at 900 °C with 488 nm laser, which shows no discernible difference between the annealed implanted samples and the unimplanted wafer.

Fig. S7 b) shows the Raman spectrum of as implanted Si, which is typical of amorphous Si where the reduced symmetry results in a Raman spectrum that is related to the phonon density of states modulated by frequency-dependent factors.[1]

Fig. S7 c) shows the peak position of the first order Raman peak as a function of annealing temperature for RTA and DC annealing and 488 and 633 nm excitation. This shows that for both excitation wavelengths, there is a red shift in the peak position for 700 and 800 °C annealing temperatures, while there is no red shift past the error of the measurement for 900 and 950 °C annealing. This red shift at the lower annealing temperature indicates the Si is under tensile stress, which was also observed in heavily B doped Si.[2] Therefore annealing above 900 °C is effective in relieves residual tensile stress from the implantation and recrystallization process.

Fig. S7 d) shows peak width of the first order Raman peak as a function of annealing temperature. For both excitation wavelengths, the FWHM increases at 700 and 800 °C annealing temperatures, while at 900 and 950 °C, the FWHM remains unchanged within the measurement error. An increase in FWHM compared to unimplanted wafer indicates a deviation from full recrystallization such as a higher dislocation density.[3] Therefore annealing above 900 °C is effective in reducing the implantation induced dislocations below a level detectable by our Raman system.

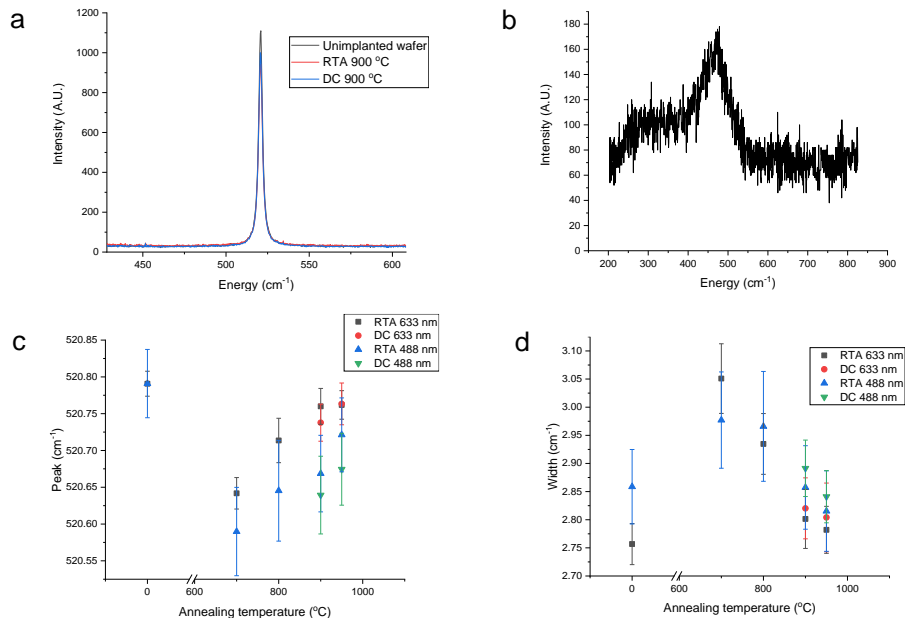

*Fig. S7 a) Primary Si Raman peak at ~520 cm$^{-1}$ of unimplanted wafer and 10$^{19}$ cm$^{-3}$ Er, 10$^{19}$ cm$^{-3}$ O co-implanted Si annealed by RTA and DC at 900 °C with 488 nm laser. b) Raman spectrum of as implanted Si with 488 nm excitation. c) peak position of the first order Raman peak as a function of annealing temperature for RTA and DC annealing and 488 and 633 nm excitation. d) peak width of the main Raman peak as a function of annealing temperature for RTA and DC annealing and 488 and 633 nm excitation. The unimplanted wafer is shown at 0 °C annealing temperature.*

### Section S4 Splitting of the $^4I_{13/2}$ manifold from Optically Modulated Magnetic Resonance and excitation measurements.

Using Optically Modulated Magnetic Resonance (OMMR), we previously measured the crystal field splitting of the $^4I_{13/2}$ manifold of Er implanted Si for the first time.[4] We performed OMMR measurements using a tuneable external cavity laser with an output power of up to ~20 mW to provide the seed signal for a C-band erbium doped fibre amplifier (EDFA) with an output power of up to ~150 mW, which was then modulated with a mechanical chopper. The EPR signal output from an EMX EPR spectrometer containing the Er:Si sample was fed into the input of a lock-in amplifier referenced to the mechanical chopper. OMMR spectra were generated by sweeping the external cavity laser wavelength and reading the lock-in signal to give a spectrum of EPR signal that has been modulated by the laser. By measuring the output spectrum of the EDFA at various external cavity laser input wavelengths, we found that at wavelengths > ~1580 nm, the output from the EDFA would include broad Amplified Spontaneous Emission (ASE) at ~1530 nm; because of this we did not include this part of the spectrum in ref [4], and we assigned the crystal field ground to the lowest energy peak indicated by black arrows at 1572 nm in the OMMR spectrum in Fig. S3.

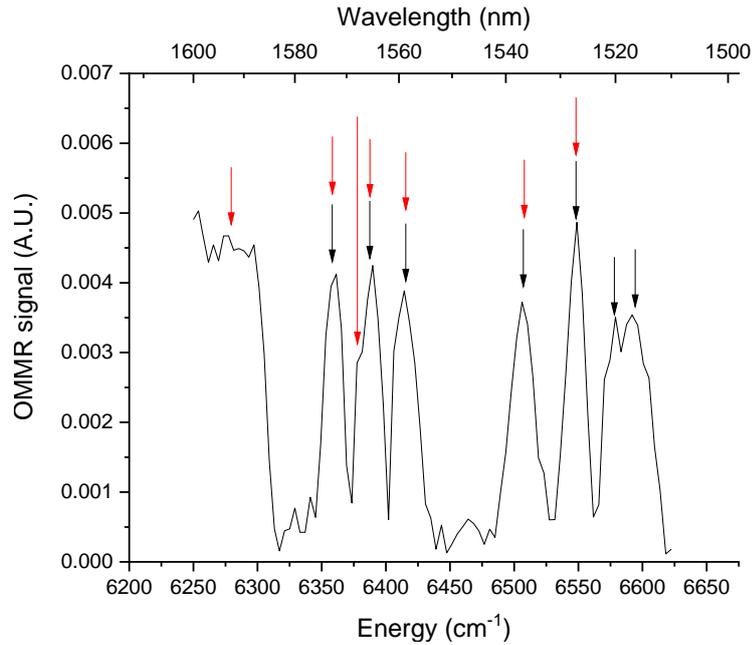

*Fig. S3 OMMR of $10^{19}$ cm$^{-3}$ Er, $10^{20}$ cm$^{-3}$ O, RTA anneal at 620 °C 3hr, 850 °C 30s at a magnetic field of 961 G. Black arrows indicate the original assignment peaks in ref [4], red arrows indicate the revised assignment of peaks based on comparison to excitation measurements of Er:Si.[5]*

By comparing our OMMR spectrum to the excitation measurements of Er:Si by Gritsch et al [5], we found that if the apparent peak in our OMMR spectrum at 1592 nm is assigned to be the $^4I_{13/2}$ crystal field ground state, there is a reasonable match to the excitation measurement by Gritsch, particularly Gritsch site A, see Fig. S4. The two highest energy peaks that are not included in the revised assignment could belong to a different Er centre. The apparent peak in our OMMR spectrum at 1592 nm is probably a genuine peak even though the EDFA output includes broad ASE because our seed laser is relatively powerful, and is still sweeping in wavelength even though ASE is being produced, this would make the peak more prominent than it really is. We showed that our OMMR spectra shifted to 227 cm$^{-1}$ lower energy than would be expected for a direct absorption measurement. This was based on the difference in energy between the lowest energy OMMR peak and the highest energy PL peak, where we chose a PL peak at 1519 nm as the transition to the $^4I_{15/2}$ ground state. We now believe that the 1519 nm peak is a hot line, and the highest energy PL transition is the standard 1536 nm peak illustrated in Fig S5. Interestingly, the revised OMMR and PL ground state peaks have the same 227 cm$^{-1}$ offset as our original assignment gave.

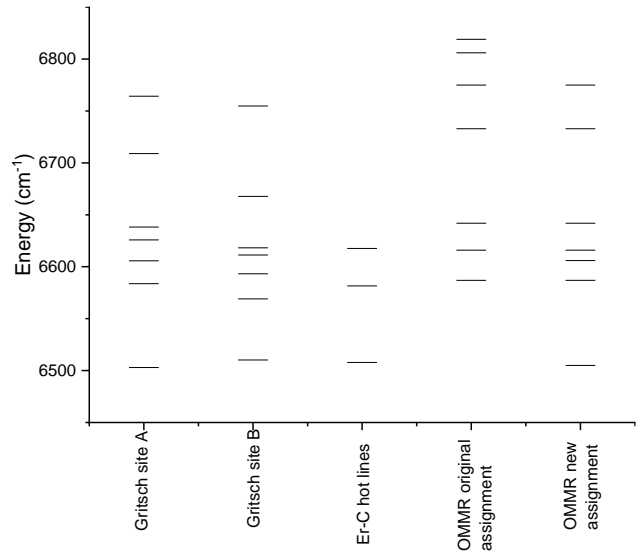

*Fig. S4 Comparison of $^4I_{13/2}$ splitting of Gritsch site A and B[5], the Er-C centre hot lines [6] and this work, OMMR original assignment [4], and the revised OMMR assignment (this work).*

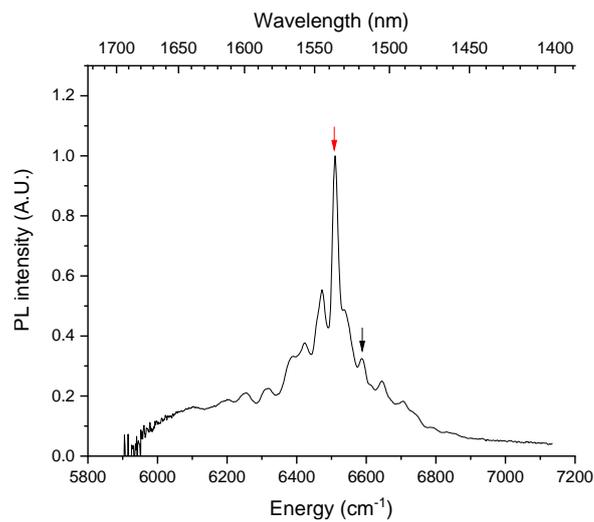

*Fig. S5 PL spectrum at 65 K of the Er:Si sample from the OMMR measurement in Fig. S3. Black arrow shows the original $^4I_{15/2}$ crystal field ground state assignment, red arrow shows the revised assignment.*

**Section S5 EPR measurements**

Fig. S9 shows the EPR measurement of the DC 950 samples at rotation angles of 0 and 30°. Where 0° rotation corresponds to the magnetic field approximately parallel to the [110] direction of the samples. Only weak isotropic resonances are visible which are associated with the cavity and residual P impurities. For comparison we show the EPR resonance of a sample optimised for EPR activity[4] with $10^{19}$ cm$^{-3}$ Er, $10^{20}$ cm$^{-3}$ O, which was annealed by RTA at 620 °C 3hr, 850 °C 50s.

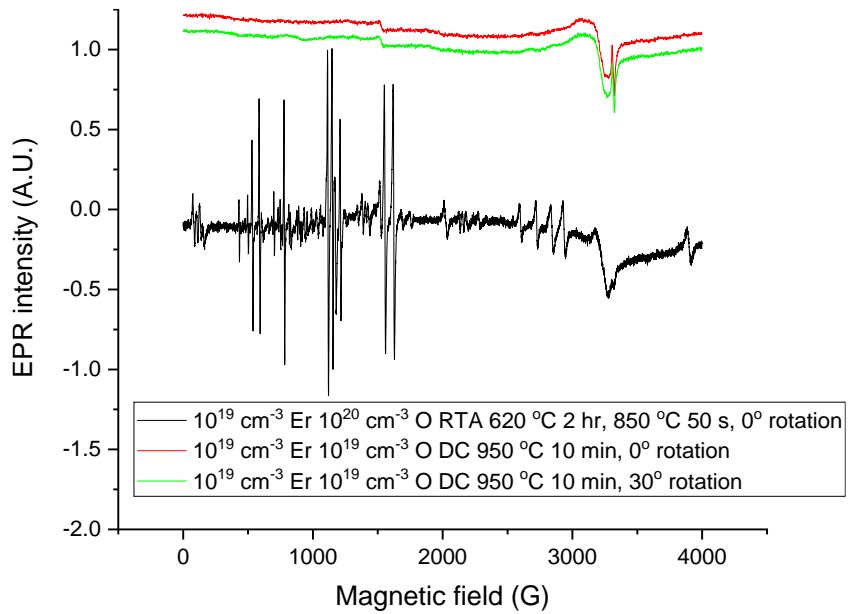

*Fig. S9 EPR of DC 950 and $10^{19}$ cm$^{-3}$ Er, $10^{20}$ cm$^{-3}$ O, RTA anneal at 620 °C 3hr, 850 °C 50s. Temperature was 5 K, MW power was 2 mW.*

**Section S6 Electrical measurements**

The resistivity measurements were performed with a standard four-point probe, the measurements were made a three different positions on the sample to obtain a mean and standard error. Fig. S8 shown that resistivity of RTA and DC samples as a function of annealing temperature. The unimplanted wafer had a resistivity of $2\times10^6$ Ωcm.

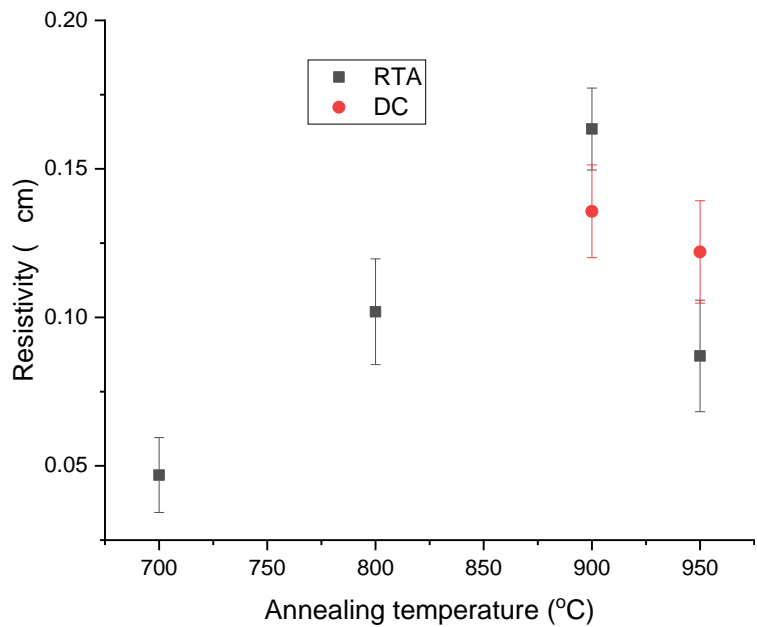

*Fig. S8 Room temperature electrical resistivity as a function of annealing temperature for RTA and DC annealing techniques. Annealing time was 10 min.*

**Supplementary References**